\documentclass[12pt]{article}
\usepackage{bezier}
\usepackage{amssymb}
\usepackage{epsfig}

\newcommand{\nc}{\newcommand}
\nc{\be}{\begin{equation}}
\nc{\ee}{\end{equation}}
\nc{\bea}{\begin{eqnarray}}
\nc{\eea}{\end{eqnarray}}

\nc{\eqn}[1]{{(\ref{#1})}}
\nc{\cA}{{\cal A}}
\nc{\cB}{{\cal B}}
\nc{\cC}{{\cal C}}
\nc{\cD}{{\cal D}}
\nc{\cE}{{\cal E}}
\nc{\cF}{{\cal F}}
\nc{\cG}{{\cal G}}
\nc{\cH}{{\cal H}}
\nc{\cI}{{\cal I}}
\nc{\cJ}{{\cal J}}
\nc{\cK}{{\cal K}}
\nc{\cL}{{\cal L}}
\nc{\cM}{{\cal M}}
\nc{\cN}{{\cal N}}
\nc{\cO}{{\cal O}}
\nc{\cP}{{\cal P}}
\nc{\cQ}{{\cal Q}}
\nc{\cR}{{\cal R}}
\nc{\cS}{{\cal S}}
\nc{\cT}{{\cal T}}
\nc{\cU}{{\cal U}}
\nc{\cV}{{\cal V}}
\nc{\cW}{{\cal W}}
\nc{\cX}{{\cal X}}
\nc{\cY}{{\cal Y}}
\nc{\cZ}{{\cal Z}}
\nc{\simo}[1]{{\stackrel{#1}{\simeq}}}
\nc{\geqo}[1]{{\stackrel{#1}{\geq}}}
\nc{\geo}[1]{{\stackrel{#1}{>}}}
\nc{\guo}[1]{{\stackrel{#1}{\succ}}}
\nc{\rbo}{\raisebox}
\nc{\RR} {\rangle \! \rangle}
\nc{\LL} {\langle \! \langle}
\nc{\rmi}[1]{{\mbox{\small #1}}}
\nc{\eq}{eq.~}
\nc{\nr}[1]{(\ref{#1})}
\nc{\ul}{\underline}
\nc{\mc}{\multicolumn}
\nc{\todo}[1]{\par\noindent{\bf $\rightarrow$ #1}}

\nc{\cu}{{\cal u}}

\title{
  \begin{flushright} {\small $\begin{array}{ l } \mbox{HD--THEP--99--29} \\
    \end{array} $}
 \end{flushright}
Renormalization of lattice gauge theories \\
with massless Ginsparg Wilson fermions}

\author{T.~Reisz$^{a,b}\,$\thanks{Supported by a Heisenberg Fellowship}
        $\;$ and
       H.~J.~Rothe$^{b}$,
         \\ \\$^a$ SPHT-Service de Physique Theorique \\
        CEA-SACLAY,\\
        F-91191 Gif-sur Yvette Cedex, France
         \\ \\$^b$ Institut
        f\"ur Theoretische Physik,\\
        Universit\"at Heidelberg, \\
        Philosophenweg 16, \\
        D-69120 Heidelberg, Germany}

\begin{document}

\maketitle

\begin{abstract}
Using functional techniques,
we prove, to all orders of perturbation theory, 
that lattice vector gauge theories with Ginsparg Wilson fermions 
are renormalizable.
For two or more massless fermions, they satisfy a flavour mixing axial
vector Ward identity. It involves a lattice specific part that is
quadratic in the vertex functional and classically irrelevant.
We show that it stays irrelevant under renormalization.
This means that in the continuum limit the (standard) chiral symmetry
becomes restored.
In particular, the flavour mixing current does not
require renormalization.
\end{abstract}

%%%%%%%%%% latex-file%%%%%%%%%%
%
% section 1 -- introduction
%
%%%%%%%%%%%%%%%%%%%%%%%%%%%%%%
%
%\section*{\label{mon.intro} I.~Introduction}
%
%%%%%%%%%%%%%%%%%%%%%%%%%%%%%%
\section{Introduction}
Recently, lattice regularization of chiral fermions
have been proposed \cite{bietenholz_wiese, neuberger} 
which circumvents the no-go theorem of Nielsen and Ninomiya 
\cite{nielsen_ninomiya}. In particular, if
the Dirac operator satisfies the Ginsparg-Wilson relation
\cite{ginsparg_wilson}
the fermion action posseses an exact chiral symmetry on
the lattice \cite{luescher_1}. 
It is free of fermion species doubling and local
in a more general sense, at least for gauge fields that are sufficiently
smooth on the scale of the UV cutoff \cite{luescher_jansen}.
In particular, $D$ decays exponentially fast,
with a decay constant proportional to the lattice cutoff.

Renormalizability of lattice gauge theories was shown in \cite{YMT}.
Crucial ingredients are the BRS symmetry of the gauge fixed
Faddeev-Popov action and of the functional measure.
According to power counting on the lattice, the 
overall UV divergencies of the vertex functional
are local lattice operators.
They are removed by adding appropriate local counterterms
to the lattice action. Linear gauge fixing that respects the discrete
lattice symmetries implies that renormalization is achieved
by renormalizing the fields and the gauge coupling constant.
The renormalized theory becomes invariant under a
(renormalized) BRS symmetry transformation.
In the proof, Wilson fermions were chosen
for the lattice realization of the Dirac operator.
This avoids the species doubling
problem and ensures that the lattice power counting theorem applies
\cite{LPT}.

Another gauge invariant realization of the fermionic action that avoids
the doubling problem are Ginsparg-Wilson fermions.
The Faddeev-Popov trick applies along the usual lines. The resulting
gauge fixed action becomes BRS invariant.
Furthermore, locality in the more general sense mentioned above and the
absense of species doubling allows one to apply the power counting theorem
again. Hence the renormalizability proof
of lattice gauge theories goes through also for the case 
of massive Ginsparg-Wilson fermions.

The renormalizability proof for massless fermions is not considerably
more involved than for massive fermions because IR and UV
singularities are properly disentangled.
For two or more Ginsparg-Wilson fermions the theory possesses an
exact chiral (flavour mixing) symmetry on the lattice.
In this paper we show that this symmetry is preserved under
renormalization.
In particular, for two massless flavours, the
renormalized vertex functional $\Gamma_R$ 
is found to satisfy the axial vector Ward identity
\bea
   && a^4 \sum_x 
    \biggl\lbrace \frac{\partial \Gamma_R}{\partial a^4 \psi(x)}
        \sigma_\alpha \gamma_5 \psi(x)
      - \overline\psi(x) \sigma_\alpha \gamma_5
          \frac{\partial \Gamma_R}{\partial a^4 \overline\psi(x)}
    \biggr\rbrace 
   \nonumber \\
   && \qquad = \; a^4 \sum_x \xi(a)
      \frac{\partial \Gamma_R}{\partial a^4 \psi(x)} 
      a \sigma_\alpha \gamma_5
      \frac{\partial \Gamma_R}{\partial a^4 \overline\psi(x)} ,
\eea
where $\{\sigma_\alpha \vert \alpha=1,2,3\}$ are the Pauli matrices
acting on the flavour components of the fermion fields.
$\xi(a)$ is such that
the right hand side is UV finite to all orders and vanishes
for $a\to 0$.
With $\Gamma_R^c = \lim_{a\to 0} \Gamma_R$ we get
\be \label{flm.chiral_cont}
   \int d^4x \; \biggl\lbrace \frac{\delta \Gamma_R^c}{\delta \psi(x)}
        \sigma_\alpha \gamma_5 \psi(x)
      - \overline\psi(x) \sigma_\alpha \gamma_5
          \frac{\delta \Gamma_R^c}{\delta \overline\psi(x)}
    \biggr\rbrace \; = \; 0 .
\ee
Hence, in the continuum limit one recovers the Ward identities
associated with the continuum axial transformations.

To avoid the technical complications of lattice QCD we will consider
in the following the renormalization of an abelian lattice gauge
theory, with two flavours of massless Ginsparg-Wilson fermions.
This model preserves the
main properties required for proving renormalizability also of non-abelian
gauge theories with massless Ginsparg-Wilson fermions.
The main inputs going into the proof are gauge or BRS invariance,
the flavour mixing axial vector Ward identity associated with an 
exact chiral symmetry and the applicability of the power counting
theorem.

%###################################################################
\section{General framework}
Consider the partition function 
\be
  Z = \int \prod_{x\in a\mathbb{Z}^4} 
    \left( d\psi(x) d\overline\psi(x) \prod_\mu dU(x;\mu)
    \right) \;
    \exp\left( - S_W(U)-S_f(U,\psi,\overline\psi)\right)
\ee
on the hypercubic lattice $a\mathbb{Z}^4$, with $a$ the lattice
spacing.
$\psi$ is a 2-flavour, Dirac spinor field,
\be
   d\psi(x) = \prod_{f=1}^2 \prod_{\alpha=1}^4
    d\psi_{f\alpha}(x)
\ee
and $dU$ is the Haar measure on $U(1)$.
$S_W(U)$ is the Wilson plaquette action
\be
   S_W(U) = \frac{1}{2g^2} \sum_{x\in a{\mathbb Z}^4} 
    \sum_{\mu\not=\nu=0}^{3}
    \biggl( 1 -  U(x;\mu) U(x+a\widehat\mu;\nu)
      U(x+a\widehat\nu;\mu)^{-1} U(x;\nu)^{-1} 
    \biggr) . 
\ee
The fermion action is given by
\be \label{flm.ferm}
  S_{f} \; = \; a^4 \sum_{x\in a {\mathbb Z}^4} 
   \overline{\psi}(x) D[U] \psi(x) .
\ee
The lattice Dirac operator $D$ is translation invariant
and is supposed to satisfies the Ginsparg Wilson relation
\be \label{flm.GW_relation}
   \gamma_5 D + D \gamma_5 \; = \; a D \gamma_5 D .
\ee
For $U(x;\mu)=\exp iagA_\mu(x)$, $D$ allows for a small $A$ expansion
\bea \label{flm.small}
  && D[\exp{iagA}](x,y) \; = \; \sum_{n\geq 0} \frac{1}{n!}
   a^{4n} \sum_{z_1,\dots ,z_n} \sum_{\mu_1,\dots ,\mu_n}
   D^{(n)}_{\mu_1\dots\mu_n}(x,y \vert z_1,\dots ,z_n)
  \nonumber \\
  && \qquad\qquad\qquad\qquad\qquad \cdot  
   A_{\mu_1}(z_1) \cdots A_{\mu_n}(z_n).
\eea
The Fourier transforms $\widetilde{D}^{(n)}$ of the
coefficient functions $D^{(n)}$ are
analytic functions about zero momentum. 

Finally, for $a\to 0$, the action becomes
\bea
   && S_W(U) + S_f(U,\psi,\overline\psi) \simeq \int d^4x \; 
    \biggl\lbrace \frac{1}{4} \sum_{\mu,\nu =0}^3 
      \left(\partial_\mu A_\nu(x)-\partial_\nu A_\mu(x)\right)^2
    \nonumber \\ 
   && \qquad  + \overline\psi(x) \sum_{\mu=0}^3 \gamma_\mu 
     ( \partial_\mu + i g A_\mu ) \psi(x)
    \biggr\rbrace .
\eea

A possible solution of (\ref{flm.GW_relation})
that satisfies these conditions is given by
\cite{neuberger,luescher_1}
\bea
   D(U) &=& \frac{1}{a}
     \left( 1 - \cD \left( \cD^+\cD \right)^{-1/2} \right) ,
   \nonumber \\
   \cD &=& 1 - a D_W ,
\eea
where $D_W$ is the Wilson Dirac operator,
\bea
   && D_W \; = \; \frac{1}{2a} \sum_{\mu=0}^3 
    \left\lbrack \left(\gamma_\mu-1\right) D_\mu^c
                + \left(\gamma_\mu+1\right) D_\mu^{c*}
    \right\rbrack
   \nonumber \\
   && \left(D_\mu^c\psi\right)(x) \; = \;
    U(x;\mu) \psi(x+a\widehat\mu) - \psi(x) ,
   \\
   && \left(D_\mu^{c*}\psi\right)(x) \; = \;
    \psi(x) -  U(x-a\widehat\mu)^{-1} \psi(x-a\widehat\mu) ,
   \nonumber
\eea
and $\widehat\mu$ the unit vector in the positive $\mu$th direction.

Measure and action are invariant under the gauge transformation
\bea
   U(x;\mu) &\to& \Lambda(x) U(x;\mu) \Lambda(x+a\widehat\mu)^{-1} 
   \nonumber \\
   \psi(x) &\to& \Lambda(x) \psi(x) ,
   \\
   \overline\psi(x) &\to& \overline\psi(x) \Lambda(x)^{-1},
   \nonumber
\eea
with all $\Lambda(x)\in U(1)$.
Furthermore, because of (\ref{flm.GW_relation})
they are invariant under the infinitesimal
global chiral transformation
\bea \label{flm.axial_t}
   \delta\psi(x) &=& i \epsilon \; \sigma_\alpha \gamma_5
     (1 - \frac{a}{2}D) \psi(x) ,
   \nonumber \\
   \delta\overline\psi(x) &=& i \epsilon \; 
     \overline\psi (1-\frac{a}{2}D)(x)
     \gamma_5 \sigma_\alpha .
\eea
Finally, we observe that the theory is invariant 
under the charge conjugation
\bea
   A_\mu(x) &\to& - A_\mu(x) ,
   \nonumber \\
   \psi(x) &\to& C \overline\psi(x)^T
   \\
   \overline\psi(x) &\to& \psi(x)^T \gamma_0^T C^{-1} \gamma_0,
   \nonumber
\eea
where the superscript $T$ denotes transposition and
$C$ the charge conjugation matrix satisfying
\be
   C^{-1} \gamma_\mu C \; = \; - \gamma_\mu^T ,
    \quad \mu=0,\dots,3 .
\ee

For the perturbative evaluation we restrict our
attention to a small
neighbourhood of the pure gauge orbit.
Here we parametrize
\be
   U(x;\mu) = \Lambda(x) \exp{\left( iag A_\mu(x) \right)}
     \Lambda(x+a\widehat\mu)^{-1},
\ee
with the $A_\mu(x)$ subject to the Lorentz gauge 
\be \label{flm.gf_lorentz}
   F(A(x)) = 
    \sum_{\mu=0}^3 \frac{1}{a} \widehat{\partial}_\mu^* A_\mu(x) \equiv 0 .
\ee
Here and in the following, $\widehat\partial$ and
$\widehat{\partial}^*$ denote the forward and backward 
lattice difference operators, respectively,
\be
  \widehat\partial_\mu f(x) = f(x+a\widehat\mu) - f(x),\;
   \widehat\partial_\mu^* f(x) = f(x) - f(x-a\widehat\mu).
\ee

Going through the standard gauge fixing and Faddeev-Popov
procedure we end up with the following generating functional
$W_0$ of the connected correlation function.
\bea \label{flm.conn_gen_functl}
   && \exp \frac{1}{\hbar} W_0(J,\eta,\overline\eta) \; = \; \cN_0 \int
    \prod_x \left( d\psi(x) d\overline\psi(x) \prod_\mu dA_\mu(x)
            \right) \;
   \nonumber \\
   && \quad \cdot \quad \exp
    \biggl\lbrace \cE_0(A,\psi,\overline\psi; g,\lambda)
           + \frac{1}{\hbar} S_c(A,\psi,\overline\psi;J,\eta,\overline\eta)
    \biggr\rbrace ,
\eea
with $W(0,0,0)=0$ and 
\be
   \cE_0 = - \frac{1}{\hbar} 
    \left( S_f(e^{iagA},\psi,\overline\psi) 
      + S_W(e^{iagA}) + S_{gf}(A) 
    \right) .
\ee
$S_{gf}$ is the gauge fixing action
\be
   S_{gf}(A) \; = \; a^4 \sum_x \frac{\lambda}{2} \;
     F(A(x))^2 .
\ee
The source term $S_c$ is given by
\bea
   && S_c(A,\psi,\overline\psi;J,\eta,\overline\eta) = a^4 \sum_x 
    \biggl\lbrace \sum_\mu J_\mu(x) A_\mu(x) 
      + \overline\eta(x)\psi(x) + \overline\psi(x)\eta(x)
    \biggr\rbrace .
\eea
Because we have chosen a linear gauge fixing function
(\ref{flm.gf_lorentz}), the ghost fields decouple and do not need to
be considered for the following considerations.
In (\ref{flm.conn_gen_functl}) we have introduced $\hbar$ mainly
as the loop counting parameter. Any connected Feynman diagram 
with $n$ loops contributes to the order of $\hbar^n$ to 
$W$ or $\Gamma$.

\section{Ward identities}

Applying an infinitesimal gauge transformation
to (\ref{flm.conn_gen_functl}),
\bea
   \delta_\omega A_\mu(x) &=& - \frac{1}{a} \widehat\partial_\mu \omega(x) 
   \nonumber \\
   \delta_\omega\psi(x) &=& i \omega(x) g\psi(x) ,
   \\
   \delta_\omega\overline\psi(x) &=& 
     - i \omega(x) g \overline\psi(x),
   \nonumber
\eea
with $\omega(x)\in{\mathbb R}$, we obtain the Ward identity
\be
   -i a^4 \sum_{\mu} \frac{1}{a}\widehat\partial_\mu^* J_\mu(x)
    + \biggl\lbrack
        g \overline\eta(x) \frac{\partial}{\partial\overline\eta(x)}
        - g \eta(x) \frac{\partial}{\partial\eta(x)}
        - i \sum_\mu \lambda \frac{1}{a^3} \widehat{\square}
        \widehat\partial_\mu^* \frac{\partial}{\partial J_\mu(x)}
      \biggr\rbrack W_0 \; = \; 0,
\ee
with $\widehat\square=\sum_\mu \widehat\partial_\mu^*\widehat\partial_\mu$.
We perform a Legendre transformation to the vertex functional 
$\Gamma_0$, i.e.~the generating
functional of the one-particle irreducible (1PI) correlation functions,
\be \label{flm.LeTr_bare}
   W_0(J,\eta, \overline\eta) = \Gamma_0(\cA,\psi,\overline\psi)
    + a^4 \sum_x 
    \biggl( \sum_\mu J_\mu(x) \cA_\mu(x) + \overline\eta(x)\psi(x)
      + \overline\psi(x)\eta(x)
    \biggr),
\ee
where
\be
   a^4 \cA_\mu(x) = \frac{\partial W_0}{\partial J_\mu(x)}, \quad
   a^4 \psi(x) = \frac{\partial W_0}{\partial\overline\eta(x)}, \quad
   a^4 \overline\psi(x) = - \frac{\partial W_0}{\partial\eta(x)},
\ee
with inversion
\be
   a^4 J_\mu(x) = - \frac{\partial \Gamma_0}{\partial \cA_\mu(x)}, \quad
   a^4 \overline\eta(x) = \frac{\partial \Gamma_0}{\partial\psi(x)}, \quad
   a^4 \eta(x) = - \frac{\partial \Gamma_0}{\partial\overline\psi(x)} .
\ee
This yields the gauge Ward identity
\be \label{flm.WI_gauge_bare}
   \cS(\Gamma_0)(x) \; = \; 0 ,
\ee
where
\bea 
   && \cS(\Gamma_0)(x) \equiv i \sum_{\mu} \frac{1}{a}\widehat\partial_\mu^* 
    \frac{\partial\Gamma_0}{\partial\cA_\mu(x)}
    + \biggl\lbrack g \overline\psi(x) 
         \frac{\partial\Gamma_0}{\partial\overline\psi(x)}
         - g \psi(x) \frac{\partial\Gamma_0}{\partial\psi(x)}
      \biggr\rbrack
   \nonumber \\
   && \qquad\qquad\qquad - i \lambda a \sum_\mu \widehat\square
         \widehat\partial_\mu^* \cA_\mu(x) .
\eea
Furthermore,
applying the chiral transformation (\ref{flm.axial_t})
to (\ref{flm.conn_gen_functl}) yields
\bea
   && 0 \; = \; 
    \prod_x \left( d\psi(x) d\overline\psi(x) \prod_\mu dA_\mu(x)
            \right) \; \exp (\frac{1}{\hbar} S_c)
   \nonumber \\
   && \qquad \cdot \; a^4 \sum_{x} 
    \biggl\lbrace \overline\eta(x) \sigma_\alpha \gamma_5
     \left( \psi(x) + \hbar \frac{a}{2} 
            \frac{\partial}{\partial a^4 \overline\psi(x)}
     \right)
   \nonumber \\
   && \qquad\qquad\quad + 
       \left( \overline\psi(x) - \hbar \frac{a}{2} 
            \frac{\partial}{\partial a^4 \psi(x)}
        \right) \gamma_5 \sigma_\alpha \eta(x)
    \biggr\rbrace
    \; \exp \cE_0
   \\
   && = \; \prod_x \left( d\psi(x) d\overline\psi(x) \prod_\mu dA_\mu(x)
              \right) \; \exp \cE_0
   \nonumber \\
   && \qquad \cdot \; a^4 \sum_{x} 
    \biggl\lbrace \overline\eta(x) \sigma_\alpha \gamma_5
     \left( \psi(x) - \hbar \frac{a}{2} 
            \frac{\partial}{\partial a^4 \overline\psi(x)}
     \right)
   \nonumber \\
   && \qquad\qquad\quad + 
       \left( \overline\psi(x) + \hbar \frac{a}{2} 
            \frac{\partial}{\partial a^4 \psi(x)}
        \right) \gamma_5 \sigma_\alpha \eta(x)
    \biggr\rbrace
    \; \exp (\frac{1}{\hbar} S_c) .
   \nonumber
\eea
In the last step we have performed a partial integration.
This expression can be written in the form
\be
   a^4 \sum_x 
    \biggl\lbrace \overline\eta(x) \sigma_\alpha \gamma_5
      \frac{\partial W_0}{\partial a^4 \overline\eta(x)}
      - \frac{\partial W_0}{\partial a^4 \eta(x)} \gamma_5
        \sigma_\alpha \eta(x)
      - \overline\eta(x) a\sigma_\alpha \gamma_5 \eta(x)
    \biggr\rbrace \; = \; 0.
\ee
Application of the Legendre tranform (\ref{flm.LeTr_bare}) yields
the flavour mixing, axial vector Ward identity
\be \label{flm.WI_axial_bare}
   \cC(\Gamma_0) \; = \; - a^4 \sum_x 
      \frac{\partial \Gamma_0}{\partial a^4 \psi(x)} 
      a \sigma_\alpha \gamma_5
      \frac{\partial \Gamma_0}{\partial a^4 \overline\psi(x)},
\ee
where we have defined
\be
   \cC(\Gamma_0) \; \equiv \; a^4 \sum_x 
    \biggl\lbrace \frac{\partial \Gamma_0}{\partial a^4 \psi(x)}
        \sigma_\alpha \gamma_5 \psi(x)
      - \overline\psi(x) \sigma_\alpha \gamma_5
          \frac{\partial \Gamma_0}{\partial a^4 \overline\psi(x)}
    \biggr\rbrace .
\ee 
The Ward identities (\ref{flm.WI_gauge_bare}) and (\ref{flm.WI_axial_bare})
express the gauge invariance and the chiral symmetry of the bare lattice
theory. 
The axial Ward identity contains a term that is bilinear
in the vertex functional and classically irrelevant, i.e.~with a
coefficient that vanishes proportional to the lattice spacing $a$.

%#########################################################################
\section{Renormalization}
We write the renormalized generating functional of connected 
correlation functions $W_R$ as
\bea
   && \exp \frac{1}{\hbar} W_R(J,\eta,\overline\eta) \; = \; \cN_R \int
    \prod_x \left( d\psi(x) d\overline\psi(x) \prod_\mu dA_\mu(x)
            \right) \;
   \nonumber \\
   && \quad \cdot \quad \exp
    \biggl\lbrace \cE_R(A,\psi,\overline\psi; g,\lambda)
           + \frac{1}{\hbar} S_c(A,\psi,\overline\psi;J,\eta,\overline\eta)
    \biggr\rbrace ,
\eea
with $W_R(0,0,0)=0$ and 
\be
   \cE_R(A,\psi,\overline\psi; g,\lambda) = 
   \cE_0(A,\psi,\overline\psi; g,\lambda) + O(\hbar^0).
\ee
The renormalized vertex functional $\Gamma_R$ is obtained from $W_R$
by the Legendre transformation analogous to (\ref{flm.LeTr_bare}).
By definition, all basic field 
($A$, $\psi$, $\overline\psi$) correlation functions exist
in the continuum limit.
In particular, this means that the connected correlation functions
\bea
   && G_{r,s,l-r-s}
       \left( (z_1,\mu_1),\dots; z_{r+1},\dots; z_{r+s+1},\dots, z_l
              \vert a
       \right)
   \nonumber \\
   && \equiv \; a^{4l} \sum_{x_1,\dots ,x_l\in a\mathbb{Z}^4}
      \prod_{i=1}^l \delta^4(z_i-x_i) \; \cdot \;
      < A_{\mu_1}(x_1) \cdots \psi(x_{r+1}) \cdots
        \overline\psi(x_l) >_c ,
\eea
with
\bea
   && < A_{\mu_1}(x_1) \cdots \psi(x_{r+1}) \cdots
        \overline\psi(x_l) >_c 
   \nonumber \\
   && = \; \hbar^l \frac{\partial}{\partial a^4 J_{\mu_1}(x_1)} \cdots
        \frac{\partial}{\partial a^4 \overline\eta(x_{r+1})} \cdots
        \frac{\partial}{\partial a^4 (-\eta(x_l))} \;
        \left. \frac{1}{\hbar} W_R \right\vert_{J=\eta=\overline\eta=0} ,
\eea
converge for $a\to 0$ as tempered distributions on 
$\mathbb{R}^{4l}$.

The renormalizability statement below is that $\cE_R$ is obtained 
from $\cE_0$ by
multiplicative renormalization of its arguments and such that both
gauge invariance and lattice chiral symmetry are preserved.

%%%%%%%%%%%%%%%%%%%%%%%%%%%%%%%%%%%%%%%%%%%%%%%%%%%%%%%%%%%%%%%%
% renormalization theorem
%%%%%%%%%%%%%%%%%%%%%%%%%%%%%%%%%%%%%%%%%%%%%%%%%%%%%%%%%%%%%%%%
{\bf Theorem}.
There exist renormalization constants 
$Z_A$, $Z_\psi$, $Z_g$ and $Z_\lambda$ such that
\be
   \cE_R(A,\psi,\overline\psi; g,\lambda) = 
   \cE_0(Z_A^{1/2} A,Z_\psi^{1/2}\psi, Z_\psi^{1/2}\overline\psi; 
         Z_g g,Z_\lambda \lambda) ,
\ee
with 
\be
   Z_A Z_g^2 \; = \; Z_A Z_\lambda \; = \; 1.
\ee
The renormalization constants are at most logarithmically divergent
as $a\to0$.

As a consequence of the theorem, the renormalized vertex functional 
$\Gamma_R$
satisfies the gauge and axial Ward identities
\bea \label{flm.WI_renormalized}
   && \cS(\Gamma_R) \; = \; 0 ,
   \nonumber \\
   && \cC(\Gamma_R) \; = \; a^4 \sum_x \xi(a)
      \frac{\partial \Gamma_R}{\partial a^4 \psi(x)} 
      a \sigma_\alpha \gamma_5
      \frac{\partial \Gamma_R}{\partial a^4 \overline\psi(x)} ,
\eea
where $\xi(a)$ is at most logarithmically divergent as $a\to 0$.
%%%%%%%%%%%%%%%%%%%%%%%%%%%%%%%%%%%%%%%%%%%%%%%%%%%%%%%%%%%%%%%%
% end renormalization theorem 
%%%%%%%%%%%%%%%%%%%%%%%%%%%%%%%%%%%%%%%%%%%%%%%%%%%%%%%%%%%%%%%%

The two independent renormalization constants
$Z_A$ and $Z_\psi$ are uniquely determined by imposing
two independent normalization conditions, e.g.~on the 
2-point functions, at non-exceptional momenta.

The statement implies that abelian lattice gauge theory with 
massless Ginsparg Wilson
fermions is renormalizable, preserving gauge and
chiral symmetry. In particular, no mass counterterm for the fermion
fields is required.
The irrelevant part of the lattice axial Ward identity gets
multiplicatively renormalized, but in such a way that it stays 
irrelevant to all orders.
This means that the continuum chiral symmetry becomes restored in
the continuum limit (that is, (\ref{flm.chiral_cont}) holds).
As a corollary, the current $j_{\mu}(x)$ associated to the
symmetry (\ref{flm.axial_t}) and constructed according to the
Poincare lemma on the lattice \cite{luescher_2}
does not require renormalization (cf.~also \cite{hasenfratz}).
For, if we write the variation of the action under a local chiral
transformation
\be
   \delta\psi(x) = i \epsilon(x) \; \sigma_\alpha \gamma_5
     [(1 - \frac{a}{2}D) \psi](x) , \quad
   \delta\overline\psi(x) = i \epsilon(x) \; 
     [\overline\psi (1-\frac{a}{2}D)](x)
     \gamma_5 \sigma_\alpha 
\ee
as
\be
   \delta \cE_R \; = \; \frac{1}{\hbar} a^4 \sum_x i \epsilon(x) 
    \left( \sum_\mu \frac{1}{a} \widehat{\partial}_\mu^* j_{\mu\alpha}(x)
    \right)
\ee
and add to the source part of the action $S_c$ a term
\be
   a^4 \sum_{x,\mu} \sum_{\alpha=1}^3 
    G_{\mu\alpha}(x) j_{\mu\alpha}(x),
\ee
the corresponding vertex functional 
$\Gamma_R^{\,\prime}(A,\psi,\overline\psi;G)$
satisfies the axial vector current Ward identity
\bea \label{flm.local_wi}
   && \sum_\mu \frac{1}{a} \widehat\partial_\mu^*
    \frac{\partial \Gamma_R^{\,\prime}}{\partial a^4 G_{\mu\alpha}(x)}
  \; = \; - 
    \biggl\lbrace \frac{\partial \Gamma_R^{\,\prime}}{\partial a^4 \psi(x)}
        \sigma_\alpha \gamma_5 \psi(x)
       - \overline\psi(x) \sigma_\alpha \gamma_5
          \frac{\partial \Gamma_R^{\,\prime}}{\partial a^4 \overline\psi(x)}
    \biggr\rbrace 
   \\
   && \qquad\qquad + \; \xi(a) 
      \frac{\partial \Gamma_R^{\,\prime}}{\partial a^4 \psi(x)} 
      a \sigma_\alpha \gamma_5
      \frac{\partial \Gamma_R^{\,\prime}}{\partial a^4 \overline\psi(x)}
      \; + \; O(G) .
   \nonumber
\eea
Because of
$\Gamma_R^{\,\prime}(A,\psi,\overline\psi;G=0)=
\Gamma_R(A,\psi,\overline\psi)$,
the right hand side of (\ref{flm.local_wi})
is UV finite to order $G^0$, and hence also the left hand side.
This implies that all correlation functions with one insertion
of the composite operator $j_\mu(x)$ are made UV finite already
by renormalizing the gauge theory.

%####################################################################
\section{Proof of the theorem}
For the proof we use the lattice power
counting theorem. Its applicability to gauge theories with
Ginsparg Wilson fermions is guaranteed by
\begin{itemize}
\item The action has the correct continuum limit.
\item Analyticity of the Fourier transforms of the coefficient functions 
$D^{(n)}$ in (\ref{flm.small}) at zero momenta.
\item The free Ginsparg Wilson propagator 
$\widetilde{D}_0(k)^{-1}$ is free of doublers.
\end{itemize}
More precisely, the last point means that $\widetilde{D}_0(k)^{-1}$ 
is of the form
\be
   \widetilde{D}_0(k)^{-1} \; = \; \frac{\cN(k)}{\cR(k)} ,
\ee
where $\cN$ and $\cR$ satisfy the following conditions.
\bea \label{flm.degree_counting}
   \lim_{a\to 0} \cN(k) &=& -i \sum_{\mu=0}^3 \gamma_\mu k_\mu, 
     \quad \overline{\rm degr}\,_{\widehat{k}} \cN(k) = 1,
   \nonumber \\
   \lim_{a\to 0} \cR(k) &=& k^2, 
     \quad \overline{\rm degr}\,_{\widehat{k}} \cR(k) = 2,
\eea
and for sufficiently small $a$ there is $K>0$ such that
\be \label{flm.prop_bound}
   \cR(k) \; \geq \; K \; \widehat{k}^2.
\ee
$\overline{\rm degr}\,_{\widehat{k}}(\cdots)$ denotes the
(UV) lattice divergence degree \cite{LPT} and
\be
   \widehat{k}^2 = \sum_{\mu=0}^3 \widehat{k}_\mu^2 \; , \qquad 
   \widehat{k}_\mu = \frac{2}{a} \sin \frac{k_\mu}{2}a .
\ee
It is straightforward to show that the propagator given in
\cite{luescher_1}, with
\bea
   && \cN(k) \; = \; \cJ(k)^{1/2} 
    \biggl\lbrack \frac{1}{a} 
      \left( \cJ(k)^{1/2}-1+\frac{a^2}{2} \widehat{k}^2 \right)
      - i \sum_\mu \gamma_\mu \widetilde{k}_\mu
    \biggr\rbrack ,
   \nonumber \\
   && \cR(k) \; = \; \widetilde{k}^2 + \frac{1}{a^2}
    \biggl( \cJ(k)^{1/2}-1+\frac{a^2}{2} \widehat{k}^2
    \biggr)^2 ,
   \\
   && \cJ(k) \; = \; 1 + \frac{a^4}{4} \sum_{\mu\not=\nu}
    \widehat{k}_\mu^2 \widehat{k}_\nu^2 ,
   \nonumber \\
   && \widetilde{k}^2 = \sum_{\mu=0}^3 \widetilde{k}_\mu^2 \; , \qquad 
    \widetilde{k}_\mu = \frac{1}{a} \sin k_\mu a ,
   \nonumber
\eea
satisfies the criteria (\ref{flm.degree_counting})
and (\ref{flm.prop_bound}) with $K=1$.

The proof of the theorem is by induction on the number of loops, 
that is on the order of $\hbar$.
To lowest order, the theorem holds, with all renormalization
constants $Z_A=Z_\psi=Z_g=Z_\lambda=1$, i.e.~all tree graph
amplitudes are UV finite. Furthermore $\xi(a)=-1$.

Let us assume that the renormalization program has been carried out
to order $n$ in $\hbar$,
with renormalized action given by
\be
   \cE_R^{(n)}(A,\psi,\overline\psi; g,\lambda) = 
   \cE_0({Z_A^{(n)}}^{1/2} A,{Z_\psi^{(n)}}^{1/2}\psi, 
          {Z_\psi^{(n)}}^{1/2}\overline\psi; 
           Z_g^{(n)} g,Z_\lambda^{(n)} \lambda) ,
\ee
and
\be \label{flm.Zren_constraints}
   Z_A^{(n)} {Z_g^{(n)}}^2 \; = \; Z_A^{(n)} Z_\lambda^{(n)} \; = \; 1,
\ee
with $Z_A^{(n)}$ and $Z_\psi^{(n)}$ at most logarithmically
UV divergent.
Hence, the vertex functional $\Gamma_R^{(n)}$ renormalized to
order $n$ satisfies
\bea
   && \cS(\Gamma_R^{(n)}) \; = \; 0 ,
   \nonumber \\
   && \cC(\Gamma_R^{(n)}) \; = \;
    a^4 \sum_x \xi^{(n)}(a) 
    \frac{\partial \Gamma_R^{(n)}}{\partial a^4 \psi(x)} 
    a \sigma_\alpha \gamma_5
    \frac{\partial \Gamma_R^{(n)}}{\partial a^4 \overline\psi(x)},
\eea
with $\xi^{(n)}(a)=-\left. Z_\psi^{(n)} \right.^{-1}$.
Considered as perturbative series in $\hbar$, $\xi^{(n)}(a)$ is at most 
logarithmically divergent.

Power counting and the properties of the lattice action listed
above ensure that all vertices derived from it have IR 
degree not less than 4 and UV degree equal to 4.
This implies that IR and UV singularities are well separated and 
that the theory is renormalizable in an IR finite way \cite{IRUV}.
The most general UV divergence of $\Gamma_R^{(n)}$ of order
$\hbar^{n+1}$, which we denote by $\Gamma_{div}$ below,
is a polynomial in the fields and their lattice derivatives,
obtained by accounting for all 1PI correlation functions
with non-negative UV divergence degree.
Taking into account the lattice symmetries, it is given by
\bea
   && \Gamma_{div} \; = \; a^4 \sum_x 
    \biggl\lbrace \delta m_\psi \overline\psi(x) \psi(x)
      + r_\psi \sum_\mu \overline\psi(x) \gamma_\mu
        \frac{1}{a} \widehat{\partial}_\mu \psi(x)
   \nonumber \\
   && \qquad + \; r_{\overline\psi A\psi} \sum_\mu 
        g \overline\psi(x) \gamma_\mu A_\mu(x) \psi(x)
      + \sum_\mu \frac{1}{2} \delta m_A^2 A_\mu(x)^2
   \nonumber \\
   && \qquad + \; \sum_{\mu\nu}
      \biggl\lbrack \frac{1}{4} r_A 
         \left( \frac{1}{a}\widehat{\partial}_\mu A_\nu(x)
               - \frac{1}{a}\widehat{\partial}_\nu A_\mu(x)
         \right)^2
   \\
   && \qquad\qquad + \; \frac{1}{2} 
           \left( r_{gf} + \delta_{\mu\nu} r_2 \right)
           \left( \frac{1}{a} \widehat{\partial}_\mu A_\mu(x) \right)
           \left( \frac{1}{a} \widehat{\partial}_\nu A_\nu(x) \right)
   \nonumber \\
   &&  \qquad\qquad + \; \frac{1}{4} 
           \left( r_{41} + \delta_{\mu\nu} r_{42} \right)
           g^2 A_\mu(x)^2 A_\nu(x)^2
      \biggr\rbrack
    \biggr\rbrace .
   \nonumber
\eea
All coefficients are $O(\hbar^{n+1})$ and all those
denoted by $r_{\bullet}$ are at most logarithmically divergent
as $a\to 0$.
Inserting this into the gauge Ward identity
\be
   \cS(\Gamma_{div})(x) 
   \; = \; O(1) 
   \quad \mbox{as}\; a\to 0
\ee
we obtain the constraints
\bea
   && r_{gf} = r_2 = r_{41} = r_{42} = \delta m_A^2 = 0 ,
   \nonumber \\
   && r_{\overline\psi A\psi} = i r_\psi .
\eea
From the axial Ward identity
\be
   \cC(\Gamma_{div}) \; = \; O(1) \quad \mbox{as}\; a\to 0
\ee
we get the further constraint $\delta m_\psi=0$.
In summary, the UV divergencies of order $\hbar^{n+1}$ 
are removed by adding to $\hbar\cE_0$ the counterterm
\bea
   && \cE_{CT} \; = \; - a^4 \sum_x 
    \biggl\lbrace  r_\psi \sum_\mu \overline\psi(x) \gamma_\mu
        \left( \frac{1}{a} \widehat{\partial}_\mu  + i g A_\mu(x) 
        \right) \psi(x)
   \nonumber \\
   && \qquad + \; \sum_{\mu\nu} \frac{1}{4} r_A 
         \left( \frac{1}{a}\widehat{\partial}_\mu A_\nu(x)
               - \frac{1}{a}\widehat{\partial}_\nu A_\mu(x)
         \right)^2
    \biggr\rbrace .
\eea
With $r_\lambda = -r_A$ and $r_g=-r_A/2$ we write this as
\bea
   && \cE_{CT} \; = \; 
     \biggl( \frac{r_A}{2} \frac{\partial}{\partial\tau_A}
            + r_g \frac{\partial}{\partial\tau_g}
            + \frac{r_\psi}{2} 
              \left( \frac{\partial}{\partial\tau_\psi}
                    + \frac{\partial}{\partial\tau_{\overline\psi}}
              \right)
            + r_\lambda \frac{\partial}{\partial\tau_\lambda}
     \biggr)
   \nonumber \\
   && \quad \cdot \; (- a^4) \sum_x 
    \biggl\lbrace  \frac{1}{4} \sum_{\mu\nu} \tau_A^2
         \left( \frac{1}{a}\widehat{\partial}_\mu A_\nu(x)
               - \frac{1}{a}\widehat{\partial}_\nu A_\mu(x)
         \right)^2
   \\
   && \qquad\quad + \; \tau_\psi\tau_{\overline\psi} \overline\psi(x) 
      \sum_\mu \gamma_\mu
      \left( \frac{1}{a} \widehat{\partial}_\mu  + i \tau_A \tau_g g A_\mu(x) 
      \right) \psi(x) 
   \nonumber \\
   && \qquad\quad + \; \frac{\lambda}{2} \tau_A^2 \tau_\lambda
      \left( \sum_\mu \frac{1}{a} \widehat{\partial}_\mu^* A_\mu(x)
      \right)^2
    \biggr\rbrace_{\tau\equiv 1} ,
   \nonumber
\eea
so that
\bea
   && \cE_R^{(n)} + \frac{1}{\hbar} \cE_{CT} \; = \;
     \cE_0( {Z_A^{(n+1)}}^{1/2} A, {Z_\psi^{(n+1)}}^{1/2} \psi, 
          {Z_\psi^{(n+1)}}^{1/2}\overline\psi; 
          Z_g^{(n+1)} g, Z_\lambda^{(n+1)} \lambda) 
   \nonumber \\
   && \qquad\qquad\qquad  + \frac{1}{\hbar} \Delta ,
\eea
with $Z_i^{(n+1)}=Z_i^{(n)}+r_i$ for
$i=A,\psi$ and with
$Z_g^{(n+1)}=\left. Z_A^{(n+1)}\right.^{-1/2}$,
$Z_\lambda^{(n+1)}=\left. Z_A^{(n+1)}\right.^{-1}$.
$\Delta$ is a local lattice operator of IR degree not
less than 4, UV degree 4, and it is UV finite and irrelevant
up to and including order $\hbar^{n+1}$.
We are thus allowed to subtract it from
$\cE_R^{(n)} + \frac{1}{\hbar} \cE_{CT}$.
This yields the action renormalized to the order $\hbar^{n+1}$, 
satisfying
\bea
   && \cE_R^{(n+1)}(A, \psi, \overline\psi ; g, \lambda) 
   \nonumber \\
   && \quad = \; 
    \cE_0( {Z_A^{(n+1)}}^{1/2} A, {Z_\psi^{(n+1)}}^{1/2} \psi, 
           {Z_\psi^{(n+1)}}^{1/2} \overline\psi; 
            Z_g^{(n+1)} g, Z_\lambda^{(n+1)} \lambda) ,
\eea
where the $Z_i^{(n+1)}$ fulfill (\ref{flm.Zren_constraints})
with $n$ replaced by $n+1$.
This closes the induction and proves the theorem.
%####################################################################

The above proof was carried out for an abelian gauge theory.
For QCD with two flavours of massless Ginsparg-Wilson fermions
the axial vector Ward identity is again of the form
(\ref{flm.WI_axial_bare}). The main complications arise 
due to the compact gauge field measure on the lattice and from the
non-linear structure of the gauge Ward identity.
The renormalizability proof proceeds along similar lines
as in \cite{YMT}, supplemented by the
axial vector Ward identity, whose
renormalized version (\ref{flm.WI_renormalized})
is obtained by following the procedure described above.

\section*{Acknowledgment}
One of us (T.R.) would like to thank Jean Zinn-Justin
for encouraging discussions.

%%%%%%%%%%%%%%%%%%%%%%%%%%%%%%
%
%  Bibliography
%
%%%%%%%%%%%%%%%%%%%%%%%%%%%%%%

%%%%%%%%%%%%%%%%%%%%%%%%%%%

%%%%%%%%%%%%%%%%%%

\end{document}